\documentstyle[epsfig,newpasp,twoside]{article}
\def\be{\begin{equation}}
\def\ee{\end{equation}}
\def\bea{\begin{eqnarray}}
\def\eea{\end{eqnarray}}

\def\etal{{\em et al.}            }

\def\Mpc{{\rm\,Mpc}}

\def\hMpc{\,h^{-1}{\rm Mpc}}

\def\spose#1{\hbox to 0pt{#1\hss}}
\def\simlt{\mathrel{\spose{\lower 3pt\hbox{$\mathchar"218$}}
     \raise 2.0pt\hbox{$\mathchar"13C$}}}
\def\simgt{\mathrel{\spose{\lower 3pt\hbox{$\mathchar"218$}}
     \raise 2.0pt\hbox{$\mathchar"13E$}}}
\def\({\left(}
\def\){\right)}
\def\[{\left[}
\def\]{\right]}
\def\<{\left\langle}
\def\>{\right\rangle}

\def\ApJ{{\em ApJ~}}
\def\MN{{\em MNRAS~}}

\def\edcomment#1{\iffalse\marginpar{\raggedright\sl#1\/}\else\relax\fi}
\reversemarginpar

\begin{document}

\title{Interpolation of Discretely-Sampled Density Fields}

\author{Will Saunders$^\ast$, Bill E. Ballinger$^\dag$}
\affil{$^\ast$IfA, University of Edinburgh, UK. $^\dag$Astrophysics, Oxford University, UK}

\begin{abstract}

We present a new technique for the interpolation of discretely-sampled non-negative scalar fields across regions of missing data. Any set of basis functions can be used, though the method is fastest when they are close to orthogonal. We show how the technique may be efficiently applied when the discrete sampling rate varies across the field. Regularisation is desirable to avoid over-fitting noisy data, and is necessary when the regions of missing data are larger than the required resolution. We present and investigate methods for such regularisation.

\end{abstract}

\section{Introduction}

Estimating the gravity and velocity fields in the Local Universe from redshift surveys requires us to estimate the galaxy density field in the parts of the sky we cannot survey, in particular behind the Milky Way. Early attempts to do this assumed a uniform density of galaxies behind the Milky Way, or attempted a crude interpolation from above and below. A great advance was the introduction of Weiner filtering techniques (Lahav \etal 1994), which combines interpolation and regularisation to provide a biased, but minimum variance, estimate of the underlying, real-space, density field. The major drawbacks are that it requires a prior estimate of the power spectrum, and also that structure not compatible with the assumption of Gaussianity is strongly suppressed; this limits its applicability to very large scales and poor resolutions.

To circumvent this, we have developed a technique where the logarithm of the underlying field is expanded as a sum of harmonics. If galaxies were actually Poisson-sampled from a lognormal underlying field, our resulting interpolated and regularised field would be the best possible estimate of the density field. This approximation is a good one (Hubble 1934, Coles and Jones 1991), and qualitatively we find our methods robust to deviations from lognormality.

\section{Mathematical Framework}

Our input data consists of real or simulated redshift surveys: that is a clustered 3D discrete distribution over a predefined solid angle, and which we model as a stochastic process drawn from an underlying continuous non-negative density field (e.g. Peebles 1980) with a spatially-dependent sampling rate which we quantify as the selection function, $\psi(r)$. Our aim is to find an expansion for the underlying density field which gives the maximum likelihood for the positions and redshifts of the galaxies in the survey, and to use this expansion to interpolate across regions with missing data. We work with the logarithm of the density field rather than the field itself, as this is both physically sensible and mathematically convenient. The basis functions can in principal be any arbitrary set of functions, but the method works faster the more nearly orthogonal they are over the volume of the survey.

So we have a set of $M$ galaxies at positions ${{\bf r}_m}$, and a set of $N$ basis functions ${f_n}$. Because of the radial symmetry inherent in all-sky surveys, we use the Bessel-Fourier basis functions, as laid out by Binney and Quinn (1991) and Heavens and Taylor (1995). These are products of spherical harmonics $Y(l,m)$ and spherical Bessel functions $B(k(l,n),r/r_{max})$, chosen to have zero derivative on the boundary at $r=r_{max}$. We also have a set of amplitudes ${a_n}$ which we wish to determine by maximising the likelihood for the observed dataset. The amplitude of the underlying density field at ${\bf r}$ is then

\be
\rho({\bf r}) = \exp{[\sum_n a_n f_n({\bf r})]}
\ee

\noindent and the likelihood for the whole survey is given by

\be 
\ln{\cal{L}} = \ln{\[\prod_m \rho({\bf r}_m)\]} = \sum_m \sum_n a_n f_n({\bf r}_m)
\ee

\noindent Clearly we need an need an integral constraint on $\rho$, otherwise the likelihood can always be increased simply by increasing the density everywhere. We demand that the total number of galaxies predicted by the density field over the unmasked region, equal the number actually observed, and we introduce this via a Lagrange multiplier. 

This gives us the main result of this paper, that 
we can find an expansion of a non-negative density field, using an arbitrary set of basis functions, which gives the maximum likelihood for a discrete distribution assumed to be stochastically sampled from the density field with known but variable sampling rate $\psi({\bf r}_m)$, by solving the $N$ equations

\be
{\partial \ln{{\cal L'}} \over \partial a_n} = \sum_m f_n({\bf r}_m)- \int f_n \psi(r) \exp{[\sum_n a_n f_n({\bf r}_m)}] dV = 0
\ee

\noindent The equations are non-linear whenever the density field itself is. In practice, we solve them by using the multidimensional Newton-Raphson technique (Press \etal 1986) - i.e. we iteratively solve them as a locally linear set of equations. This involves finding and inverting the information matrix 

\be
{{\partial^2 \ln{{\cal L'}} \over \partial a_n \partial a_{n'}}} = \int f_n f_{n'} \psi({\bf r}_m) \exp[{\sum_n a_n f_n({\bf r}_m)}] dV  
\ee

The more nearly linear the field, and the more nearly orthogonal the basis functions, the faster the method works. Typically 10 iterations are sufficient.
The solution ${a_n}$ of the equations (9) defines a density field that fills the entire volume, even if the data itself does not do so. As long as the regions of missing data are smaller than the resolution of the expansion, the solution is perfectly stable and gives a natural interpolation across these regions. Indeed, if the field is actually lognormal, the set ${a_n}$ are independent Gaussian deviates, and hence the reconstruction in the gaps is also lognormal with identical first and second moments to the regions with data - that is, they are statistically identical.

\subsection{Radial Rescaling} 

For predicting e.g. the gravity dipole on the Local Group, we would like to go out to a distance of $200 \hMpc$ or so while keeping a resolution of a $\Mpc$ or so nearby, requiring at first sight millions of modes. However, our method necessarily involves repeated matrix inversions with as many rows and columns as basis functions, and this limits us to a few thousand harmonics. We have thus made use of the suggestion in Fisher \etal (1995), of first transforming the radial coordinate of the survey so as to make the selection function unity. That is, we define a new radial coordinate $R$, where 

\be 
\frac{1}{3} R(r)^3 = \int_0^r r^2 \psi(r') dr'
\ee

Note that this means that the resolution element becomes increasingly prolate with distance. We have chosen our maximum values of $l$ and $n$ such that the resolution is roughly isotropic at the median distance of the survey. In the rescaled survey, the shot noise is the same everywhere. 

\subsection{Regularisation}
When we try and push to higher wavenumbers, the reconstruction unsurprisingly becomes unstable in regions of missing data, and some form of regularisation is needed. An obvious solution is to add a penalty term to the likelihood involving the amplitude of the harmonics, derived from their {\em a priori} distribution based on the power spectrum. However, we have not taken this approach because firstly, we would like to avoid external input; secondly our radial transform makes this calculation rather ugly. Instead, we have formulated an analogous penalty function suggested by Andrew Hamilton, based on the real as opposed to transformed density field: at each step in the iteration, we estimate the rms variation in $\ln{\rho}$ in radial shells, and fit the log of this with a cubic polynomial, giving a smooth $\sigma(r)$. We then add a penalty term to the likelihood of 

\be
\Delta = - K \int_V ({\ln{\rho}\over \sigma(r)})^2 dV
\ee

\noindent where $K= N/2V$ gives filtering analogous to Weiner filtering, while our preferred value   $K= N/4V$ gives a power-preserving filter.

\subsection{Error estimates}

From the information matrix, we can easily determine error maps for the reconstruction. The inverse of this matrix is the covariance matrix $C_{ij}$. The error in $\ln{\rho}$ at any position is then given by

\be
{\rm Var}(\ln{\rho}) = \sum_i \sum_j C_{ij} f_i f_j
\ee

\noindent which is straightforward though laborious to calculate.

\subsection{Tests on simulations}

Figure 1 shows the results of applying the algorithm to a CDM simulation with $\Gamma = 0.25$, first given the whole $4\pi$ coverage, and then restricted to the BTP and then the PSCz area. In each case, we have used harmonics up to $l=8$, $n=12$. Away from the Plane, the reconstructions are virtually identical. The PSCz reconstruction in the Plane is in general good, except behind the Galactic Centre where the mask is widest - e.g. (20,0,2000), (20,0,6000), (350,-10,8000). The BTP reconstruction is much better.

\begin{figure}
\centerline{\epsfig{figure=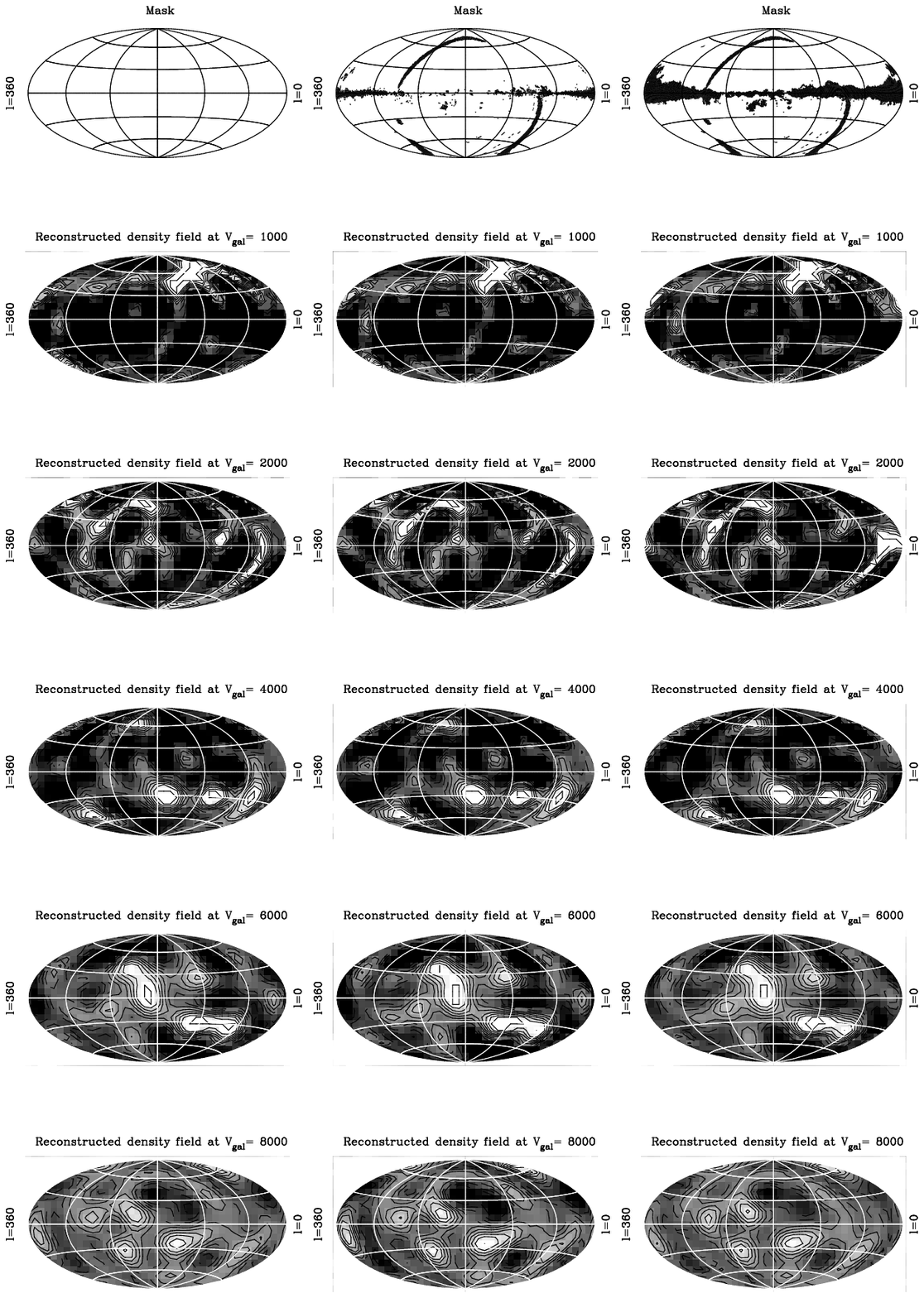,width=13.4cm}}
Figure 1. Mask and reconstructed density maps for a CDM $\Gamma = 0.25$ simulation, with all-sky coverage (first column), and with the BTP (second column) and PSCz (third column) mask.
\end{figure}

\end{document}